\documentclass[aps,prl,twocolumn,superscriptaddress,showpacs]{revtex4}
\usepackage{amsfonts,amssymb,amsmath,latexsym,epsfig,times} 
\begin{document}
[Phys. Rev. Lett. {\bf 93}, 098701 (2004)]\\ \\

\title{Cascade control and defense in complex networks}

\author{Adilson E. Motter}
\email{motter@mpipks-dresden.mpg.de}
\affiliation{Max Planck Institute for the Physics of Complex Systems,
N\"othnitzer Strasse 38, 01187 Dresden, Germany}

\date{\today}

\begin{abstract}

Complex networks with heterogeneous distribution of loads may undergo a global
cascade of overload failures when highly loaded nodes or edges are removed due
to attacks or failures.  Since a small attack or failure has the potential to
trigger a global cascade, a fundamental question regards the possible strategies
of defense to prevent the cascade from propagating through the entire network.
Here we introduce and investigate a costless strategy of defense based on a
selective further removal of nodes and edges, right after the initial attack or
failure.  This intentional removal of network elements is shown to drastically
reduce the size of the cascade.  

\end{abstract}


\maketitle

The problem of attacks on complex networks \cite{rv}
has attracted a great deal of attention in recent years
\cite{AJB:2000,CEAH:2000,CNSW:2000,CEAH:2001,ML:2002}.
The key factor prompting this research is the observation that in many growing
networks some nodes evolve to become much more important
than others \cite{BA:1999,ASBS:2000}.  From a global perspective, the important
nodes are those whose removal may either cause the network to fragment or
severely limit the communication between the other nodes.  
A prime paradigm in this context is that of the
scale-free networks (SFNs), which 
are characterized by an algebraic distribution of
degrees (number of edges per node) \cite{BA:1999}.  While robust against random
removals \cite{AJB:2000,CEAH:2000,PV:2001}, SFNs with realistic
scaling exponents \cite{rv} are likely to fragment into small clusters if a
critical fraction of most connected nodes is removed
\cite{AJB:2000,CNSW:2000,CEAH:2001}.  This happens because of the significant
number of edges removed along with the highly connected nodes
\cite{CEAH:2001}.

But in networks where the flow of a physical quantity is important, such as
power grids and computer networks, the fragmentation of the network may be
triggered by the removal of one or very few nodes.  Indeed, although the removal
of few nodes has little effect on the connectivity of the network, these
removals may trigger a cascade of subsequent failures which can in turn switch
off or disconnect most of the other nodes of the network, such as in the August
14, 2003 event in the northeastern U.S.~power transmission network and in recent
blackouts in western Europe.  A number of important aspects of cascading
failures in complex networks have been discussed in the literature, including
disturbances in power transmission systems \cite{SCL:2000}, the origin
of rare events \cite{Watts:2002}, the effect of network growth \cite{HK:2002},
cascades triggered by intentional attacks \cite{ML:2002}, avalanche size
distributions \cite{MGP:2002}, and congestion instabilities
\cite{MPVV:2003} (see also
Refs.~\cite{comm,AAN:2004}).  In particular, a
simple model of cascades of {\it overload} failures has been introduced
\cite{ML:2002}.  The removal of an even small fraction of highly loaded nodes (due to
either attack or failure) has been shown to trigger global cascades of overload
failures in SFNs and other networks with heterogeneous distribution of
loads.  This calls for an investigation on possible strategies of
defense to prevent the cascade from propagating through the entire network.

In this Letter we study a general method to reduce the size of cascades of
overload failures triggered by attacks on or failures of a small fraction of
nodes (hereafter referred to as {\it the initial attack}).  One such cascade can
be divided in two parts:  (I) the initial attack, whereby a fraction of nodes is
removed; and (II) the propagation of the cascade, where another fraction of
nodes is removed due to the subsequent overload failures.  The size of the
cascade is measured in terms of the ratio $G=N'/N$, where $N$ and $N'$ are the
initial (before (I)) and final (after (II)) number of nodes in the largest
connected component, respectively.  The method of defense introduced here
consists of costless modifications to the network structure after (I) but before
(II).  In real networks, (I) and (II) are separated in time but this time
interval is usually much shorter than the time scale in which the network
evolves.  Therefore, it is reasonable to consider that no edge or node can be
rewired or added to the system after the initial attack because any of these
operations would involve extra costs.  Accordingly, we assume that, after the
initial attack, the only operations allowed in order to reduce the size of the
cascade are the {\it intentional removals} of nodes and edges.  The expression
``intentional removals'' (IRs)
means removals performed after (I) and before (II).
The trick point here is that these removals in general reduce even more the
final number of nodes in the largest connected component.  We show, however,
that the IR of {\it carefully chosen} nodes and edges can in
fact constitute an efficient strategy of defense.  Our main result is that the
size of the cascade can be drastically reduced with the IRs of
nodes having {\it small load} and edges having {\it large excess of load} (to be
defined below).  Even though any removal always increases the immediate damage
on the network, the resulting $G$ is in this case significantly larger (as
compared to the case without defense) because {\it these} IRs
strongly suppress the propagation of the cascade.

For concreteness, we 
consider the model of overload failures introduced in
Ref.~\cite{ML:2002}, which is defined as follows.  For a given network ${\cal
W}$, we assume that at each time step one unit of a physical quantity (hereafter
called {\it packet}) is sent from node $i$ to node $j$, for every ordered pair
of nodes $(i,j)$ belonging to the same connected component of ${\cal W}$.  We
assume that the packet is transmitted along the shortest paths connecting nodes
$i$ and $j$.  If there is more than one shortest path connecting two given
nodes, the packet is divided evenly at each branching point.  The load $L_k$ on
a node $k$ is the total amount of packets passing through that node per unit of
time \cite{GKK:2001}.  Let $S_k$ denote the connected component of node $k$ and
$L_k^{(i,j)}$ denote the contribution of the ordered pair $(i,j)$ to the load on
$k$.  The load on node $k$ is then \cite{GKK:2003}
\begin{equation}
L_k = \sum_{i,j} L_k^{(i,j)},
\label{0}
\end{equation}
where the sum is over all pairs of nodes in $S_k$.  Each node $k$ is assigned to
have a finite capacity $C_k$.  The node operates in a free-flow regime if $L_k
\leq C_k$, otherwise the node is assumed to fail and is removed from the
network.  Now consider that ${\cal W}= {\cal W}(0)$ is an initially connected
network.  The initial load $L_k=L_k(0)$ is given by Eq.~({\ref{0}}) with
$S_k={\cal W}(0)$.  The capacity $C_k$ of node $k$ is assumed \cite{ML:2002} to
be proportional to the initial load $L_k(0)$:
\begin{equation}
C_k=\lambda L_k (0), \;\;\; k=1,2,...N,
\label{2}
\end{equation}
where $\lambda\geq 1$ is the tolerance parameter and $N$ is the initial number
of nodes in the network.
The cascade can be regarded as a
step-by-step process where nodes can be removed at each time step.  We start
with ${\cal W}={\cal W}(0)$ at time $0$.  The condition $\lambda\geq 1$
guarantees that no node of ${\cal W}(0)$ is overloaded, i.e., $L_k(0)\leq C_k$
$\forall k$.  We assume that an initial attack is performed at time 1, whereby a
fraction $p$ of nodes is removed from ${\cal W}(0)$ and the resulting network is
denoted by ${\cal W}(1)$.  These removals lead to a global redistribution of
loads among the remaining nodes in the network.  The updated load $L_k(1)$ on a
particular node $k$ of ${\cal W}(1)$ may then become larger than the capacity
$C_k$.  All the overloaded nodes are removed simultaneously from ${\cal W}(1)$
and the resulting network is denoted by ${\cal W}(2)$.  This leads to a new
redistribution of loads and subsequent overloads may occur.  The overloaded
nodes are removed and the resulting network is denoted by ${\cal W}(3)$, and so
on.  Let ${\cal W}(n)$ denote the updated network at time $n$.  This cascading
process stops only when, for a certain $n=n'$, the updated load satisfies
$L_k(n')\leq C_k$ for all the nodes $k$ of ${\cal W}(n')$.  The ratio $G$
introduced above as a measure of the size of the cascade is defined in terms of
the number $N'$ of nodes in the largest connected component of ${\cal W}(n')$.

Here we focus on global cascades  [i.e., $1-G = O(1)$] triggered by initial attacks on a small
fraction $p$ of {\it most} loaded nodes.  To be specific, we consider a random
model of SFNs \cite{NSW:2001}, where the degree ${\kappa}_i\ge {\kappa}_0$ of
each node $i$ is chosen at random according to the probability distribution
$P({\kappa})\propto {\kappa}^{-\gamma}$, for a given scaling exponent $\gamma$ and
constant integer ${\kappa}_0$.  The degree ${\kappa}_i$ can be regarded as the
number of ``half edges'' emerging from node $i$.  A network ${\cal W}(0)$ is then
generated by randomly connecting half edges to form edges, prohibiting self- and
repeated edges \cite{sf}.  The initial attack and the propagation of the cascade
correspond to (I)  ${\cal W}(0) \rightarrow {\cal W}(1)$ and (II)  ${\cal
W}(1) \rightarrow {\cal W}(2) \cdots\rightarrow {\cal W}(n')$, respectively.
Our method of defense consists of an intermediate step  ${\cal W}(1)
\rightarrow {\cal W}(1)$ in between (I) and (II), whereby IRs
of nodes and/or edges are performed on ${\cal W}(1)$.  The network ${\cal W}(1)$
is redefined to incorporate these removals.  

We first consider the IR of nodes.  In what follows, all the
quantities refer to the initial network ${\cal W}(0)$ (unless explicitly
mentioned otherwise) and `$(0)$' is omitted for simplicity.  We recall that
$L_k^{(i,j)}$ denotes the contribution of the ordered pair of nodes $(i,j)$ to
the load on node $k$.  The total contribution of the (unordered) pair $i$ and
$j$ to the load on the network is $\sum_k L_k^{(i,j)} + \sum_k
L_k^{(j,i)}=2(D_{ij}+1)$, where $D_{ij}$ is the shortest path length between
nodes $i$ and $j$.  The factor $2$ comes from the fact that one packet is sent
from node $i$ to node $j$ and another is sent from node $j$ to node $i$. 
We consider that half of this amount is {\it
generated} by node $i$ while the other half is generated by node $j$.  Then the
total load generated by node $i$ is
\begin{equation}
L_i^g= \sum_j (D_{ij}+1)= (\bar{D}_i+1)(N-1),
\label{3}
\end{equation}
where $\bar{D}_i$ is the average shortest path length from node $i$ to all the
others.  The average load on each node is $\bar{L}=\sum_i L_i/N=\sum_i L_i^g/N =
(\bar{D}+1)(N-1)$, where $\bar{D}=\sum_i \bar{D}_i/N$ is the average shortest
path length between any two nodes.  Nodes whose load $L_i$ is much larger
than $L_i^g$ contribute much more to handling than to generating load.  These
are the most important nodes for the network to operate.  The removal of one
such node may cause overloads on a number of other nodes.  More precisely, if a
node $i$ is removed from the network, the total load on the remaining nodes
increases by at least the amount $L_i-2L_i^g$, unless the removal of node $i$
divides the network into more than one connected component.  Along with the
observation that nodes having large $L_i$ tend to have small $L_i^g$
\cite{corr}, this provides a rationale for attack strategies based on the
removal of highly loaded nodes.  On the other hand, nodes whose load $L_i$ is
smaller than $L_i^g$ generate more load than they handle.  This observation is
our starting point to argue that the size of the cascade can be drastically
reduced with the IR of a certain fraction $f$ of nodes
according to any of the following strategies:  (1) nodes with smallest $\Delta_i
\equiv L_i-L_i^g$ are removed first; (2) nodes with smallest {\it closeness
centrality} $\bar{D}_i^{-1}$ are removed first; (3) nodes with smallest load
$L_i$ are removed first; (4) nodes with smallest degree $\kappa_i$ are removed
first.  The IRs are performed on ${\cal W}(1)$, but for
simplicity we assume that $\Delta_i$, $\bar{D}_i$, $L_i$, and $\kappa_i$
correspond to the initial values computed on ${\cal W}(0)$.  In random SFNs, all
these quantities are strongly correlated \cite{corr}:  the quantities $L^g_i$
and $L_i$ are negatively correlated, while the quantities $\Delta_i$,
$\bar{D}_i^{-1}$, $L_i$, and $\kappa_i$ are positively correlated.  Therefore we
only need to justify one of the strategies of defense (1)-(4).  Consider then
strategy (1).

Strategy (1) consists in removing a fraction $f$ of nodes with most negative
$\Delta_i$.  In order to reduce the size of a cascade triggered by an initial
attack, these IRs have to satisfy two conditions.  The first
condition is to reduce the load on the remaining nodes in the network.  Because
$\Delta_i$ is negatively correlated with $\bar{D}_i$, the removal of nodes with
$\Delta_i<0$ tends to reduce the average shortest path length $\bar{D}$ between
the nodes that remain connected to the largest component.  The average load $\bar{L}$
on a node of this component is proportional to both $\bar{D}+1$ and the number of other nodes
in the same component, and is therefore reduced with strategy (1).  
This argument is based on the initial network ${\cal W}(0)$ but, for large
random networks, the same is expected to hold true on ${\cal W}(1)$ as well.
The second condition is that the fragmentation caused by the 
IRs must be smaller than that otherwise caused by the cascade itself, i.e.
$\tilde{G}>G^o$, where $\tilde{G}$ and $G^o$ denote the fraction of nodes remaining in
the largest connected component right after the IRs and after the cascade without
defense, respectively.  Because $\Delta_i$ is positively correlated with the degree
$\kappa_i$, the nodes removed according to strategy (1) tend to be the least
connected nodes in the network.  It is well known that most of the (unremoved)
nodes remain in a single connected component when any fraction of least
connected nodes is removed \cite{AJB:2000,CEAH:2000}.  More specifically,
$\tilde{G}$ is expected to decrease linearly as $\tilde{G}\approx 1-f$ with the
fraction $f$ of nodes removed according to strategy (1).  (We assume that the
fraction $p$ of nodes removed in the initial attack is small enough so that we
can neglect the contribution of the initial attack to $\tilde{G}$.)  Therefore,
the second condition is satisfied even for relatively large $f$ insofar as
$f<1-G^o$.  The ratio $G$ for the cascade with defense is $G^*=\tilde{G} -\Delta
\tilde{G}$, where $\Delta\tilde{G}$ is due to the propagation of the cascade.
Gathering all these, because $\tilde{G}$ decreases slowly with increasing $f$,
there should be a certain $f<1-G^o$ for which $\tilde{G}$ is large (as compared
to $G^o$) and the average load on the remaining nodes is sufficiently reduced so
that the propagation of the cascade is strongly suppressed and $\Delta
\tilde{G}$ is small.  Therefore strategy (1) is expected to be effective in
reducing the size of global cascades.  Because of the correlations mentioned
above, the same is expected for the strategies (2)-(4).

On the other hand, the network is expected to be sensitive to the removal of
nodes with large $\Delta_i$ and, due to the correlations \cite{corr}, to the
removal of nodes with large $\bar{D}_i^{-1}$, $L_i$, and $\kappa_i$ as well.
All these quantities can be regarded as measures of {\it centrality}.  Therefore
our first result could be stated as follows:  while the removal of the most
central nodes of ${\cal W}(0)$ can trigger global cascades, the removal of the
least central nodes of ${\cal W}(1)$ can drastically reduce the size of these
cascades.

\begin{figure}[t]
\begin{center}
\epsfig{figure=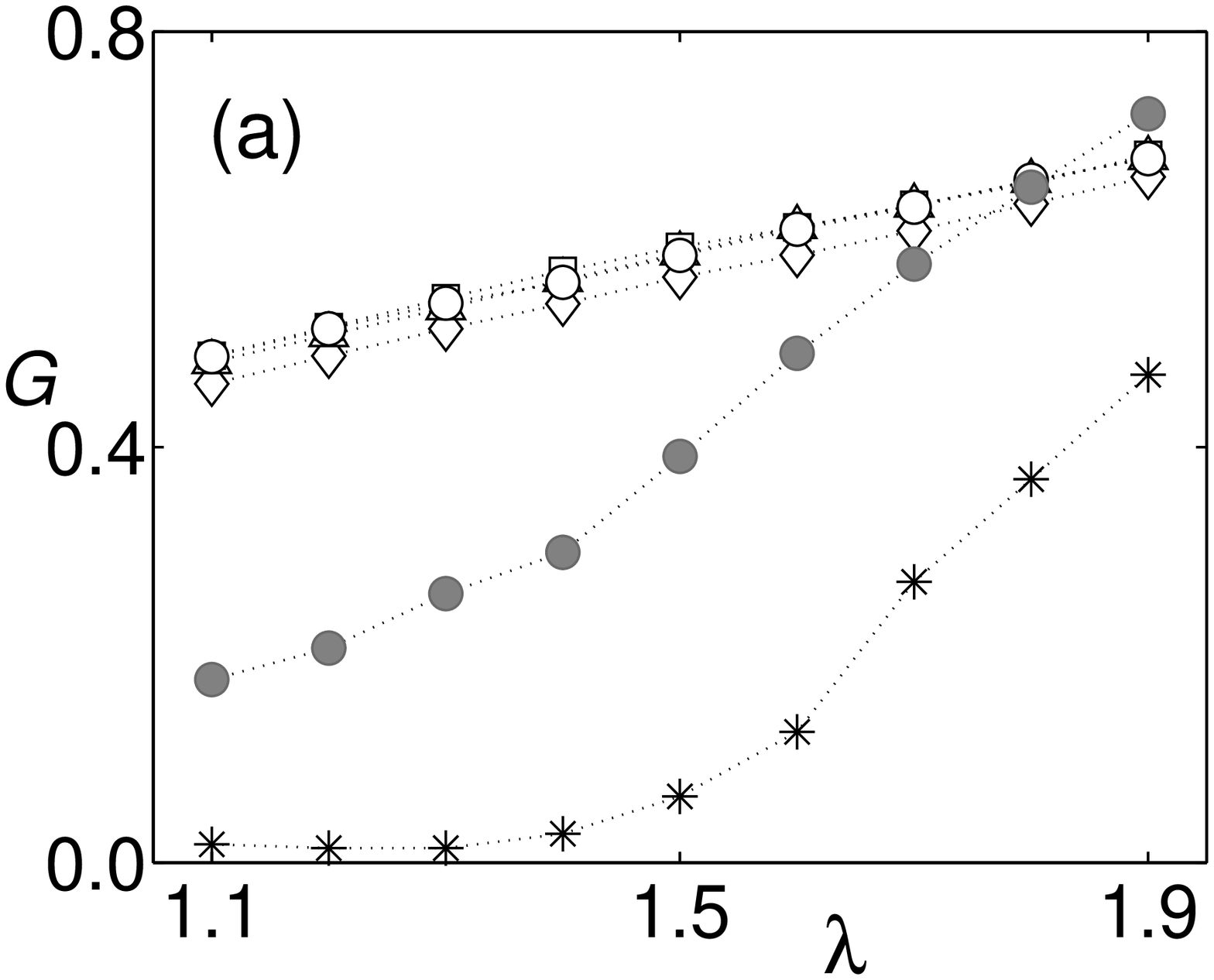,width=4.1cm}
\epsfig{figure=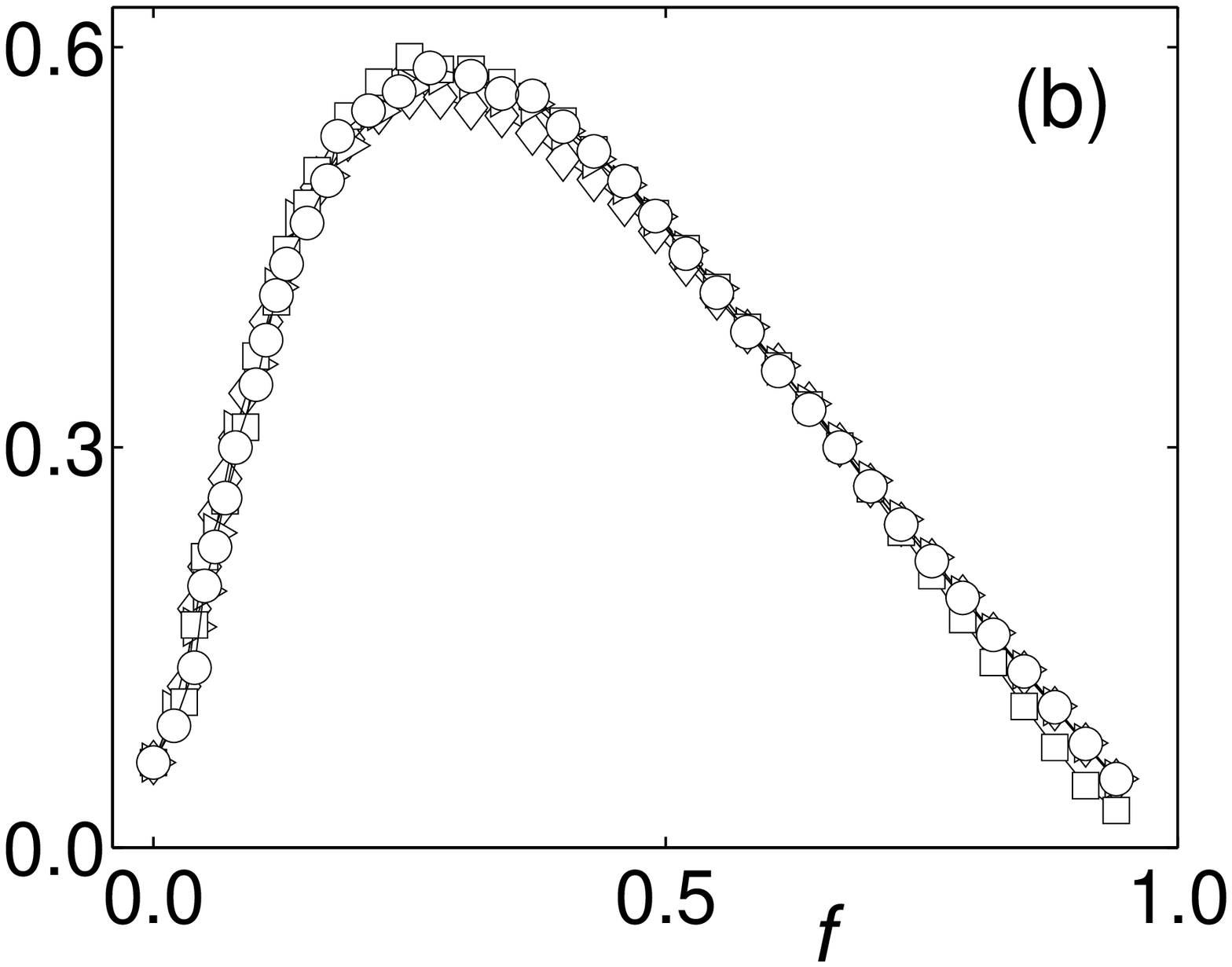,width=4.0cm}
\caption
{(a) Ratio $G$ as a function of the tolerance parameter $\lambda$.  Stars
correspond to attacks without defense, while open circles, squares, triangles,
and diamonds correspond to the IR of nodes according to the strategies of
defense (1)-(4), respectively.  (b) Ratio $G$ as a function of the fraction $f$
of nodes intentionally removed according to each of the strategies (1)-(4), for
$\lambda=1.5$. 
Solid circles in (a) correspond to the IR of edges for $\xi=2\lambda$.  Each curve corresponds to an average
over 20 independent realizations of the network for $\gamma=3.0$, $\kappa_0= 2$,
$N=5000$, and $p=0.001$.}
\label{fig1}
\end{center}
\end{figure}

Now we present numerical verification of our result concerning the IR of nodes.
We consider random SFNs with scaling exponent
$\gamma$ and initial attacks on a small fraction $p\ll 1$ of most loaded nodes.
Strong evidence for our result is presented in Fig.~\ref{fig1}(a), where
we show the ratio $G$ as a function of the
tolerance parameter $\lambda$ for $\gamma=3.0$ and $p=0.001$. Without defense,
this initial attack on only $0.1 \%$ of the nodes triggers global cascades
even for relatively large values of the tolerance parameter $\lambda$ [Fig.~\ref{fig1}(a), stars].
However, 
the ratio $G$
is shown to be 
significantly larger when a suitable fraction of nodes is
intentionally removed according to any of the strategies of defense (1)-(4)
[Fig.~\ref{fig1}(a), open symbols]. 
For example, for 
$\lambda=1.5$, we have $G\approx 0.6$ with defense and $G\approx 0.06$ without it.  
A similar improvement is observed for other values of the scaling exponent $\gamma$.
As a function of the fraction $f$ of IRs, the ratio $G$
displays a 
well-defined maximum, as shown in Fig.~\ref{fig1}(b) for $\lambda=1.5$. 
When $f$ is large, the propagation of the cascade is strongly suppressed and nearly all
the damage is caused by the IRs, i.e., $G$ is approximately $1 - f$.
This explains the linear behavior of $G$ for $f > 0.4$ in Fig.~\ref{fig1}(b).
When $f$ is small, most of the damage is caused by the cascade itself.
The maximum of $G$ lies in a region of
intermediate $f$ where the propagation of the cascade
is significantly suppressed and the damage caused by
the IRs is relatively small.
The results presented in Fig.~\ref{fig1}(a)
correspond to this maximum.  The almost perfect agreement between the
different strategies of defense in Figs.~\ref{fig1}(a) and \ref{fig1}(b) is due to the
strong correlations between loads, path lengths, and degrees in random SFNs.

We now turn to the IR of edges.  We argue that the size of the
cascade can be drastically reduced with the IR of edges not
necessarily connected to the nodes removed in the IR of nodes.
In analogy to the load on nodes, we define the load $L^{e}_{ij}$ on an edge
between nodes $i$ and $j$ as the total amount of packets passing through that
edge per unit of time.  The load $L_i$ on node $i$ can be expressed in terms of
the load on edges as
\begin{equation}
L_i=\frac{1}{2}\sum_{j \sim i}L^{e}_{ij} + (N_i-1),
\label{4} 
\end{equation}
where the sum is over all the edges directly connected to node $i$ and $N_i$ is
the number of nodes in the corresponding connected component $S_i$.  Now
consider the network ${\cal W}(1)$ right after the initial attack.  From 
Eq.~(\ref{4}), for the load $L_i(1)$ on a node $i$ to be larger than the capacity
$C_i=\lambda L_i(0)$, the load on at least one of the edges of node $i$ must
exceed the initial load by a factor larger than $\lambda -1$, i.e.,
$L^{e}_{ij}(1)>\lambda L^{e}_{ij}(0)$ for some edge connected to $i$.  We say
that this edge has {\it large excess of load}.  The IR of one
such edge is expected to reduce the load on node $i$ (except for exceptional
cases where the load on $i$ remains unchanged).  From this follows that the
IR of edges having large excess of load can be used to reduce
the size of the cascade.  However, for these removals to be effective we need to
take into account that the removal of an edge may cause overloads elsewhere.
As a simple strategy that takes this into account, we propose to remove edges
from ${\cal W}(1)$ according to an {\it auxiliary} cascade of overload failures
{\it on edges}.  This is not to be taken as a real cascade but instead as a
criterion to remove edges.  Specifically, we consider the following:  ({\it i})
each edge has an imaginary capacity $C_{ij}^{e}=\xi L^{e}_{ij}(0)$, where
$\xi\ge 1$ is a tunable parameter; and ({\it ii}) the resulting cascade on edges
consists of a step-by-step process where all the edges satisfying $L^{e}_{ij}(1)
> C_{ij}^{e}$ are removed simultaneously from the network and ${\cal W}(1)$ is
redefined at each step to incorporate these removals.  Note that the cascade on
edges, and hence the IR of edges, takes place after the initial
attack (I) and before the propagation of the cascade on nodes (II).  Because the
condition $L^{e}_{ij}(1)>\lambda L^{e}_{ij}(0)$ is necessary but not sufficient
for node $i$ to be overloaded, we expect this strategy to be effective for suitable
$\xi>\lambda$.  Without making effort to optimize the result, in our
computations we set $\xi=2\lambda$.  The size of the cascades is significantly
reduced with this strategy, as shown in Fig.~\ref{fig1}(a) (solid circles)
for random SFNs with scaling exponent $\gamma=3.0$.  For example, for
$\lambda=1.5$, the IR of edges increases $G$ by a factor of
$6$.  The effectiveness of this strategy for intermediate values of $\xi$
expresses a trade-off between the suppression of the cascade on nodes (enhanced
for smaller $\xi$) and the fragmentation of the network due to the 
IR of edges (reduced for larger $\xi$).

In summary, we have introduced a general method to reduce the size of cascades
of overload failures triggered by attacks on or failures of highly loaded nodes.
The method is based on the IR of nodes and edges before the propagation of the
cascade.  We have shown that the size of the cascade can be drastically reduced
with the IR of nodes having small load and/or edges having large excess of load.
Here we have focused on SFNs and we emphasize that similar results are expected
for other networks with heterogeneous distribution of loads.  In the
model considered above the packets are transmitted along the shortest paths
between nodes, but a heterogeneous distribution of load is also expected for
many models with local routing.  Even in a model where the packets are routed
randomly, packets tend to visit highly connected nodes much more often than
other nodes \cite{NR:2003}.  If, on one hand, this heterogeneity in the load
distribution makes the system vulnerable to cascading failures, on the other
hand, it makes the removal of poorly loaded nodes and excessively loaded edges
an effective method to reduce the size of the cascades.\\ 

The author thanks Eduardo  Gu\'eron for illuminating discussions.


\begin{references}


\bibitem{rv}
R. Albert and A.-L. Barab\'{a}si, Rev. Mod. Phys. {\bf 74}, 47 (2002);
S.N. Dorogovtsev and J.F.F. Mendes, Adv. Phys. {\bf 51}, 1079 (2002);
M.E.J. Newman, SIAM Rev. {\bf 45}, 167 (2003). 



\bibitem{AJB:2000}
R. Albert {\it et al.}, Nature (London) {\bf 406}, 378 (2000).

\bibitem{CEAH:2000}
R. Cohen {\it et al.}, Phys. Rev. Lett. {\bf 85}, 4626 (2000).

\bibitem{CNSW:2000}
D.S. Callaway {\it et al.}, Phys. Rev. Lett. {\bf 85}, 5468 (2000).

\bibitem{CEAH:2001}
R. Cohen {\it et al.}, Phys. Rev. Lett. {\bf 86}, 3682 (2001); {\bf 87}, 219802 (2001);
S.N. Dorogovtsev and J.F.F. Mendes, {\it ibid} {\bf 87}, 219801 (2001). 


\bibitem{ML:2002}
A.E. Motter and Y.-C. Lai, Phys. Rev. E {\bf 66}, 065102 (2002).



\bibitem{BA:1999}
A.-L. Barab\'{a}si and R. Albert, Science {\bf 286}, 509 (1999).

\bibitem{ASBS:2000}
L.A.N. Amaral {\it et al.},
Proc. Natl. Acad. Sci. U.S.A. {\bf 97}, 11149 (2000).


\bibitem{PV:2001}
A related issue regards the absence of an epidemic threshold in SFNs
[R. Pastor-Satorras and A. Vespignani, Phys. Rev. Lett. {\bf 86}, 3200 (2001)].



\bibitem{SCL:2000} 
M.L. Sachtjen {\it et al.}, Phys. Rev. E {\bf 61},  4877 (2000);
B.A. Carreras {\it et al.}, Chaos {\bf 12}, 985 (2002).

\bibitem{Watts:2002}
D.J. Watts, Proc. Natl. Acad. Sci. U.S.A. {\bf 99}, 5766 (2002).

\bibitem{HK:2002}
P. Holme and B.J. Kim, Phys. Rev. E {\bf 65}, 066109 (2002).

\bibitem{MGP:2002} 
Y. Moreno {\it et al.}, Europhys. Lett. {\bf 58}, 630 (2002);
K.-I. Goh {\it et al.}, Phys. Rev. Lett. {\bf 91}, 148701 (2003).

\bibitem{MPVV:2003}
Y. Moreno {\it et al.}, Europhys. Lett. {\bf 62}, 292 (2003).



\bibitem{comm}
R. Guimer\`a {\it et al.}, Phys. Rev. Lett. {\bf 89}, 248701 (2002);
P. Holme, Adv. Complex Syst. {\bf 6}, 163 (2003);
P. Crucitti {\it et al.}, Phys. Rev. E {\bf 69}, 045104 (2004).

\bibitem{AAN:2004}
R. Albert {\it et al.}, Phys. Rev. E {\bf 69}, 025103 (2004).


\bibitem{GKK:2001}
K.-I. Goh {\it et al.}, Phys. Rev. Lett. {\bf 87}, 278701 (2001);
{\bf 91}, 189804 (2003);
M. Barth\'{e}lemy,  {\it ibid} {\bf 91}, 189803 (2003).

\bibitem{GKK:2003}
K.-I. Goh {\it et al.},  Physica A {\bf 318}, 72 (2003).



\bibitem{NSW:2001}
M.E.J. Newman {\it et al.},  Phys. Rev. E {\bf 64}, 026118 (2001).

\bibitem{sf} We take $\kappa_0\geq 2$ and impose the constraints that $\kappa_i$
be smaller than the number of nodes, $\sum_i \kappa_i$ be even, and
${\cal W}(0)$ be connected.


\bibitem{corr} Using the generating function formalism \cite{NSW:2001}, we have
derived the following approximate expression for the average value of
$\bar{D}_i$ at nodes with degree $\kappa$:  $\langle\bar{D}_i\rangle\approx
1+\ln(N/\kappa)/\ln z$, where $z$ is the ratio between the average number of
second and first neighbors of a node.  In addition, as shown in
Ref.~\cite{GKK:2001}, the average load on nodes with degree $\kappa$ scales as
$\langle L_i\rangle\sim \kappa^{\eta}$, where $\eta>1$ for SFNs
with scaling exponents in a wide interval around $\gamma =3.0$.  Thus, the
quantities $L^g_i=(\bar{D}_i+1)(N-1)$ and $L_i$ are negatively correlated, while
the quantities $\Delta_i=L_i-L_i^g$, $\bar{D}_i^{-1}$, $L_i$, and $\kappa_i$ are
positively correlated.



\bibitem{NR:2003}
L.A. Adamic {\it et al.},  Phys. Rev. E {\bf 64}, 046135 (2001);
J.D. Noh and H. Rieger, Phys. Rev. Lett. {\bf 92}, 118701 (2004).



\end{references}
\end{document}